\documentclass[onecolumn]{emulateapj}
\usepackage{graphicx}
\usepackage{natbib}

\renewcommand{\baselinestretch}{2}

\long\def\symbolfootnotetext[#1]#2{\begingroup%
\def\thefootnote{\fnsymbol{footnote}}\footnotetext[#1]{#2}\endgroup} 

\shorttitle{}
\shortauthors{\"Oberg et al.}

\begin{document}

\title{A cold complex chemistry toward the low-mass protostar B1-b: evidence for complex molecule production in ices$^*$\symbolfootnotetext[1]{Based on observations carried out with the IRAM 30-m telescope. IRAM is supported by INSU/CNRS (France), MPG (Germany) and IGN (Spain).}
}

\author{ }
\affil{ }

\author{Karin I. \"Oberg\altaffilmark{1,2}}
\affil{Harvard-Smithsonian Center for Astrophysics, MS 42, 60 Garden St, Cambridge, MA 02138, USA.}

\author{Sandrine Bottinelli}
\affil{Centre d'Etude Spatiale des Rayonnements, 9 avenue du Colonel Roche, BP 4346, 31028 Toulouse Cedex 4, France.}

\author{Jes K. J{\o}rgensen}
\affil{Centre for Star and Planet Formation, Natural History Museum of Denmark, University of Copenhagen, {\O}ster Voldgade 5-7, 1350 Copenhagen K., Denmark}

\author{Ewine F. van Dishoeck\altaffilmark{3}}
\affil{Leiden Observatory, Leiden Sterrewacht, P.O. Box 9513, 2300 RA Leiden, The Netherlands}

\altaffiltext{1}{Leiden Observatory, Leiden Sterrewacht, P.O. Box 9513, 2300 RA Leiden, The Netherlands}
\altaffiltext{2}{Hubble fellow.}
\altaffiltext{3}{Max-Planck Institute f{\"u}r Extraterrestrische Physik, Giessenbachstr. 1, 85748 Garching, Germany}

\begin{abstract}
Gas-phase complex organic molecules have been detected toward a range of high- and low-mass star-forming regions at abundances which cannot be explained by any known gas-phase chemistry. Recent laboratory experiments show that UV irradiation of CH$_3$OH-rich ices may be an important mechanism for producing complex molecules and releasing them into the gas-phase. To test this ice formation scenario we mapped the B1-b dust core and nearby protostar in CH$_3$OH gas using the IRAM 30m telescope to identify locations of efficient non-thermal ice desorption. We find three CH$_3$OH abundance peaks tracing two outflows and a quiescent region on the side of the core facing the protostar. The CH$_3$OH gas has a rotational temperature of $\sim$10~K at all locations. The quiescent CH$_3$OH abundance peak and one outflow position were searched for complex molecules. Narrow, 0.6-0.8 km s$^{-1}$ wide, HCOOCH$_3$ and CH$_3$CHO lines originating in cold gas are clearly detected, CH$_3$OCH$_3$ is tentatively detected and C$_2$H$_5$OH and HOCH$_2$CHO are undetected toward the quiescent core, while no complex molecular lines were found toward the outflow. The core abundances with respect to CH$_3$OH are $\sim$2.3\% and 1.1\% for HCOOCH$_3$ and CH$_3$CHO, respectively, and the upper limits are 0.7--1.1\%, which is similar to most other low-mass sources. The observed complex molecule characteristics toward B1-b and the pre-dominance of HCO-bearing species  suggest a cold ice (below 25~K, the sublimation temperature of CO) formation pathway followed by non-thermal desorption through e.g. UV photons traveling through outflow cavities. The observed complex gas composition together with the lack of any evidence of warm gas-phase chemistry provide clear evidence of efficient complex molecule formation in cold interstellar ices.

\end{abstract}


\section{Introduction}

Complex organic molecules have been detected toward a range of astrophysical environments, including low-mass protostars \citep{vanDishoeck95, Cazaux03, Bottinelli07}; however, the origins of these complex molecules as well as their fates are uncertain. Commonly-suggested formation routes for the detected molecules include various gas-phase reactions starting with thermally evaporated CH$_3$OH ice, atom-addition reactions on dust grains, and UV- and cosmic ray-induced chemistry in the icy grain mantles that form during the pre-stellar stages \citep[][for a review]{Charnley92, Nomura04, Herbst09}. The focus is currently on an ice formation pathway \citep[e.g.][]{Garrod06, Garrod08} because of the failures of gas phase chemistry to explain the observed abundances of some of the most common complex molecules toward low-mass protostars, especially HCOOCH$_3$. 

Recent experiments on the photochemistry of CH$_3$OH-rich ices have shown that 1) UV irradiation of CH$_3$OH ices at 20--70~K result in the production of large amounts of the complex molecules observed around protostars and 2) the chemistry has a product branching ratio which is temperature and ice composition dependent \citep{Oberg09d}. A key result is that HCOOCH$_3$ and other HCO-X ices are only abundantly produced in CO:CH$_3$OH mixtures. A similar result has been reported for proton-bombarded CH$_3$OH and CH$_3$OH:CO ices \citep{Bennett07b}. CO:CH$_3$OH ice mixtures are probably common since CH$_3$OH forms from hydrogenation of CO \citep[e.g.][]{Watanabe03,Cuppen09}. CO evaporates at 17--25~K on astrophysical time scales \citep{Bisschop06}, and thus HCO-X ices will mainly form in cold, UV-exposed ices. At higher temperatures, closer to the protostar, UV irradiation of the remaining pure CH$_3$OH ice will instead favor the production of C$_2$H$_5$OH, CH$_3$OCH$_3$ and (CH$_2$OH)$_2$. 

In light of this proposed formation scenario, protostars rich in CH$_3$OH ice are natural targets when searching for complex molecule sources. The low-mass protostar B1-b is such a source. From previous SCUBA and 3 mm continuum maps, B1-b consists of two cores B1-bN and B1-bS, separated by 20'' \citep{Hirano99}. The cores are similar with $T_{\rm dust}\sim18$~K, $M=1.6-1.8\:M_{\odot}$ and $L_{\rm bol}=2.6-3.1\:L_{\odot}$. The {\it Spitzer Space Telescope} only observed one protostar, to the south-west of the B1-bS dust core at 03:33:20.34, +31:07:21.4 (J2000), but it was still named B1-b -- to avoid confusion it will be referred to as the `protostar' position. From the {\it Spitzer} $c2d$ (From Molecular Cores to Planet Forming Disks) ice survey the B1-b protostar has a CH$_3$OH ice abundance of 11\% with respect to H$_2$O ice \citep{Boogert08}, corresponding to $\sim5\times10^{-6}$ with respect to H$_2$. The fractional abundance of H$^{13}$CO$^+$ decreases by a factor of 5 toward the SCUBA core, indicative of CO freeze-out \citep{Hirano99}. A HCO-dominated complex chemistry is thus expected where the ice is exposed to UV radiation.

The B1-b region is complicated by a number of outflows, which may enhance the UV field through shocks and open cavities through which stellar UV  photons can escape \citep{Jorgensen06, Walawender09, Hiramatsu10}. The outflow of most interest to this study runs in the south-west direction from B1-b protostar, though it is not known whether B1-b is its source or just happens to lie in its path. The outflow is not observed north-east of the B1-b protostar, where the B1-b SCUBA core is situated. It is therefore unclear whether the outflow terminates at the protostar or continues into the core, hidden from view. If it does penetrate into the SCUBA core, this may increase the UV flux in the region by orders of magnitude, enhancing both the UV photochemistry and photodesorption of ices.

The latter is important, since once formed, the complex molecules must be (partly) released into the gas phase to be observable at millimeter wavelengths. The second reason for targeting B1-b is the detection of a large CH$_3$OH gas phase column density of $\sim$2.3--2.5$\times10^{14}$ cm$^{-2}$ toward the protostar \citep{Oberg09a,Hiramatsu10} and thus ice evaporation. The narrow line width and low excitation temperature of the observed CH$_3$OH suggest non-thermal ice evaporation -- the abundances are consistent with UV photodesorption. Assuming that complex ice chemistry products desorb non-thermally at an efficiency similar to CH$_3$OH ($\sim$10$^{-3}$ molecules per incident UV photon \citep{Oberg09d}),  the protostellar environment should contain a cold gas-phase fingerprint of the complex ice composition; B1-b may offer the first possibility to observe the earliest stages of complex molecule formation, untainted by either warm ice chemistry or gas-phase processing.

Addressing these predictions, we present IRAM 30-m telescope observations of a $80''\times80''$ CH$_3$OH map of the B1-b protostellar envelope and nearby dust cores, followed by a search for HCOOCH$_3$, CH$_3$CHO, CH$_3$OCH$_3$, C$_2$H$_5$OH and HOCH$_2$CHO toward two of the identified CH$_3$OH peaks. The CH$_3$OH gas abundances and the complex abundances and upper limits with respect to CH$_3$OH are then discussed in terms of different ice desorption mechanisms and complex molecule formation scenarios. The results are finally compared with complex molecule observations toward other low-mass protostars and outflows.

\section{Observations and data reduction}

The B1-b protostellar envelope and the nearby SCUBA cores were mapped in three CH$_3$OH transitions in February 2009 using the HEterodyne Receiver Array HERA \citep{Schuster04} on the Institut de RadioAstronomie Millim\'etrique (IRAM) 30-m telescope together with the VESPA autocorrelator backend (Tables \ref{tab:obs} and \ref{tab:ch3oh}). The $80''\times80''$ map is centered on R.A 03:33:21.3 and Dec 31:07:36.7 (J2000), the estimated midpoint between B1b-S and B1b-N \citep{Hirano99}. VESPA was used with a 320 kHz channel spacing resulting in a velocity resolution of 0.78 km s$^{-1}$ (channel spacing $\delta$V of $\sim$0.39 km s$^{-1}$) at 242 GHz. Observations were carried out through two separate undersampled (8'' spacing) raster maps using beam-switching with a throw of 200'' in azimuth. The two maps were off-set by +4'', + 4'' and the combination of the two maps thus results in a single over-Nyquist-sampled map -- the half power beamwidth of the 30-m telescope at 242 GHz is $\sim$10''. The rms is  $\sim$50~mK in the first map and $\sim$130~mK in the second map in a 0.39 km s$^{-1}$ channel. The average rms over the entire map was estimated to $\sim$90~mK. The system temperature varied between 600 and 1200~K. 

The second set of observations aimed at detecting complex molecules were carried out in June 2009 with the IRAM 30-m telescope, using the new EMIR receiver (Table \ref{tab:obs}). The positions used for pointing correspond to the CH$_3$OH maximum toward the center of the SCUBA core at 03:33:20.80, 31:07:40.0 (J2000), termed the 'core' position, and a local maximum at 03:33:21.90, 31:07:22.0 (J2000) termed the 'outflow' position. The targeted lines are listed in Table \ref{tab:comp} and the frequencies are taken from the JPL molecular database and the Cologne Database for Molecular Spectroscopy \citep{Muller01}. The focus is on lines with low excitation energies (15--100~K) because of the previously discovered low excitation temperature of CH$_3$OH toward the same source. The chosen settings contain at least two lines for HCOOCH$_3$, CH$_3$CHO, CH$_3$OCH$_3$ and C$_2$H$_5$OH (and additional HOCH$_2$CHO and C$_2$H$_5$CN lines) to allow us to constrain the excitation temperatures of any detected species. The observations were carried out using three different receiver settings, each including a combination of the EMIR 90 and 150 GHz receivers (Table \ref{tab:obs}). At these wavelengths, the beam sizes are $\sim$27 and 19'', respectively. All three settings were used toward the core position, and two toward the outflow position. Each receiver was connected to a unit of the autocorrelator, with a spectral resolution of 40~MHz and a bandwidth of 120~MHz, equivalent to an unsmoothed velocity resolution of $\sim$0.1 km s$^{-1}$. Typical system temperatures were 100-150 K. All observations were carried out using wobbler switching with a 140'' throw (100'' in azimuth and 100'' in elevation). 

For both sets of observations, pointing was checked every $\sim$2 hr on J0316+413 with a typical pointing accuracy of $\sim$2--3''. All intensities reported in this paper are expressed in units of main-beam brightness temperature, which were converted from antenna temperatures in paKo using reported main beam and forward efficiencies (B$_{\rm eff}$ and F$_{\rm eff}$) of $\sim$75 and 95\% with the E90 receiver, $\sim$69 and 93\% with E150 receiver and a main beam efficiency of 52\% at 2 mm for the HERA receiver. The rms in mK are reported in Table \ref{tab:obs}.

The data were reduced with the CLASS program, part of the GILDAS software package (see http://www.iram.fr/IRAMFR/GILDAS). Linear (first-order) baselines were determined from velocity ranges without emission features, and then subtracted from the spectra. Some velocity channels showed spikes, which were replaced by interpolating the closest two good channels in the map, while no correction was made to the complex molecule spectra -- there were no spikes close to any expected line positions and thus no correction was necessary.  The complex organics observation were smoothed to velocity resolutions of  0.4-0.8 km s$^{-1}$ toward the core position and 2-4 km km s$^{-1}$ toward the outflow position respectively to achieve a maximum signal-to-noise while still barely resolving the expected lines.
\section{Results}

\subsection{CH$_3$OH}

\begin{figure}[htp]
\plotone{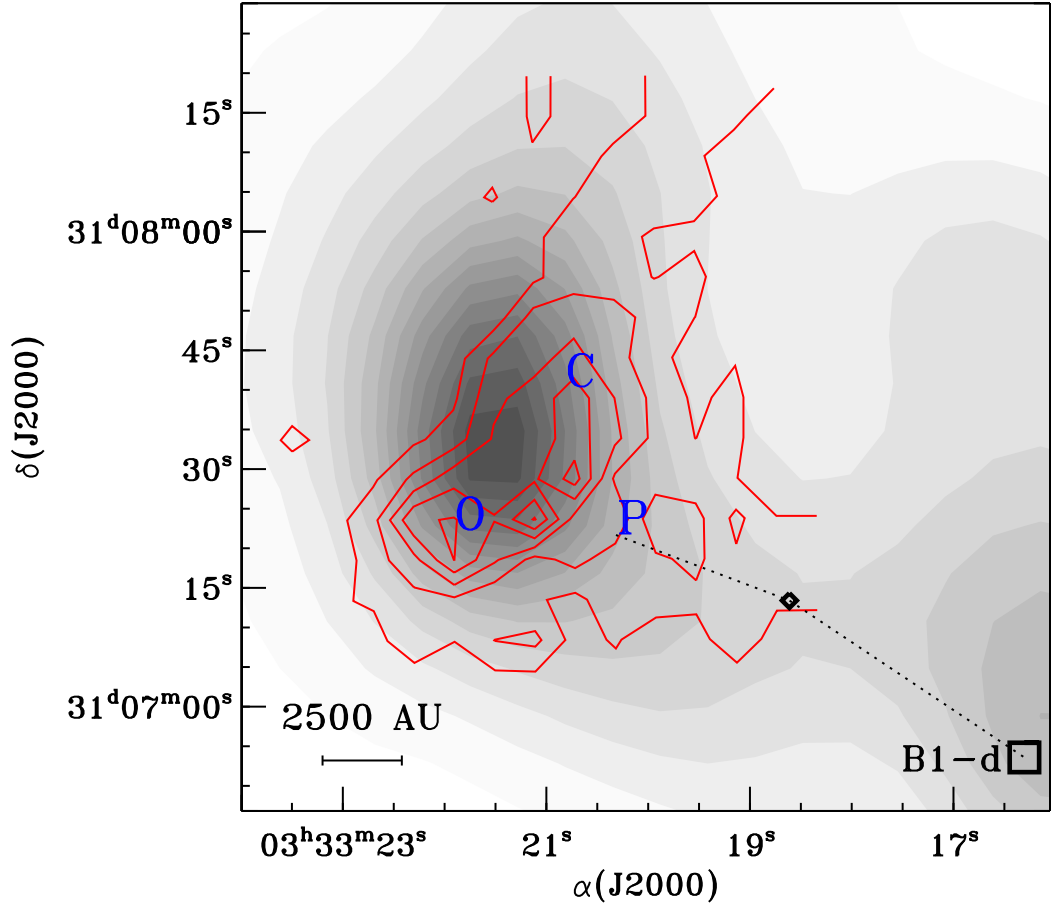}
\caption{An integrated CH$_3$OH-A$^+$ 5$_{0 5}$-- 4$_{0 1}$ map toward the B1-b region (contours) plotted together with a SCUBA 850 $\mu$m map (gray scale). The `P' marks the `protostar' position, the `C' the `core' position and `O' the `outflow' position. The CH$_3$OH contours are $\sim$2$\sigma$ i.e. 0.6~K~km~s$^{-1}$. The map also shows the position of nearby protostar B1-d and the main outflow knot between the two protostars (diamond) as well as possible outflow  trajectories from either B1-b or B1-d to the knot. \label{fig:int}}
\end{figure}

\begin{figure}[htp]
\plotone{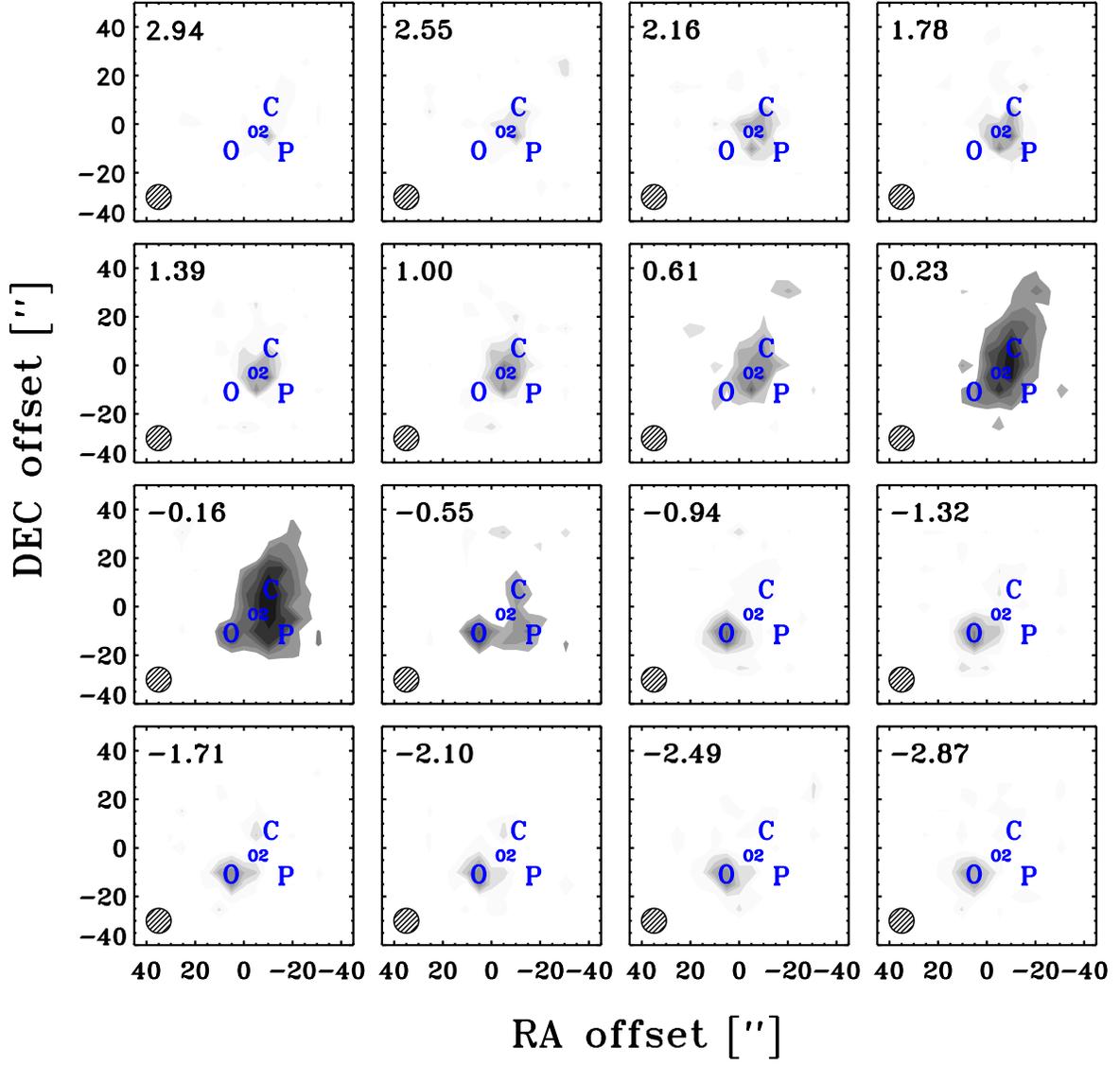}
\caption{A channel map of the CH$_3$OH-A$^+$ 5$_{0 5}$-- 4$_{0 1}$ line, centered on the B1-b SCUBA dust core. The contours are $\sim$1$\sigma$ i.e. 0.1 K and the V$_{\rm lsr}$ of each channel is listed in the upper left corners. The complex molecule observations were pointed toward the CH$_3$OH core (C) seen clearest around 0 km s$^{-1}$  and the south-west outflow (O) visible at $<$-0.55 km s$^{-1}$. The second perpendicular outflow is seen in the $>$1.0 km s$^{-1}$ channels and is labeled with O2. \label{fig:map}}
\end{figure}

Figure \ref{fig:int} shows the integrated CH$_3$OH map superimposed on a SCUBA dust map of the B1-b region. The dust map contains one elongated core, rather than the two sources observed by \citet{Hirano99}. The SCUBA core in the remainder of the paper refers to the dust core in Fig. \ref{fig:int}, while the `core' position refers to the nearby CH$_3$OH maximum probed for complex molecules. The dust and molecular maps share some features, such as the N-S elongation and the 'tail' toward the south-west, but the centers of the two maps are offset -- the CH$_3$OH column density peaks in between the SCUBA core and the `protostar' suggestive of ice desorption on the west side of the dust core due to irradiation from the protostar traveling through outflow cavities, as discussed in more detail below. 
A velocity channel map of the same CH$_3$OH line shows that in addition to the quiescent, elongated CH$_3$OH emission that overlaps with the western half of the dust emission, there are at least two high velocity components, suggestive of two perpendicular outflows which are either directed along the line of sight or which are too young to have any resolved component in the plane of the sky (Fig. \ref{fig:map}). The two outflow positions are almost coincidental with B1-bS as reported by \citet{Hirano99} and may thus be evidence for a very young protostellar object embedded in the B1-bS core.

\begin{figure}[htp]
\plotone{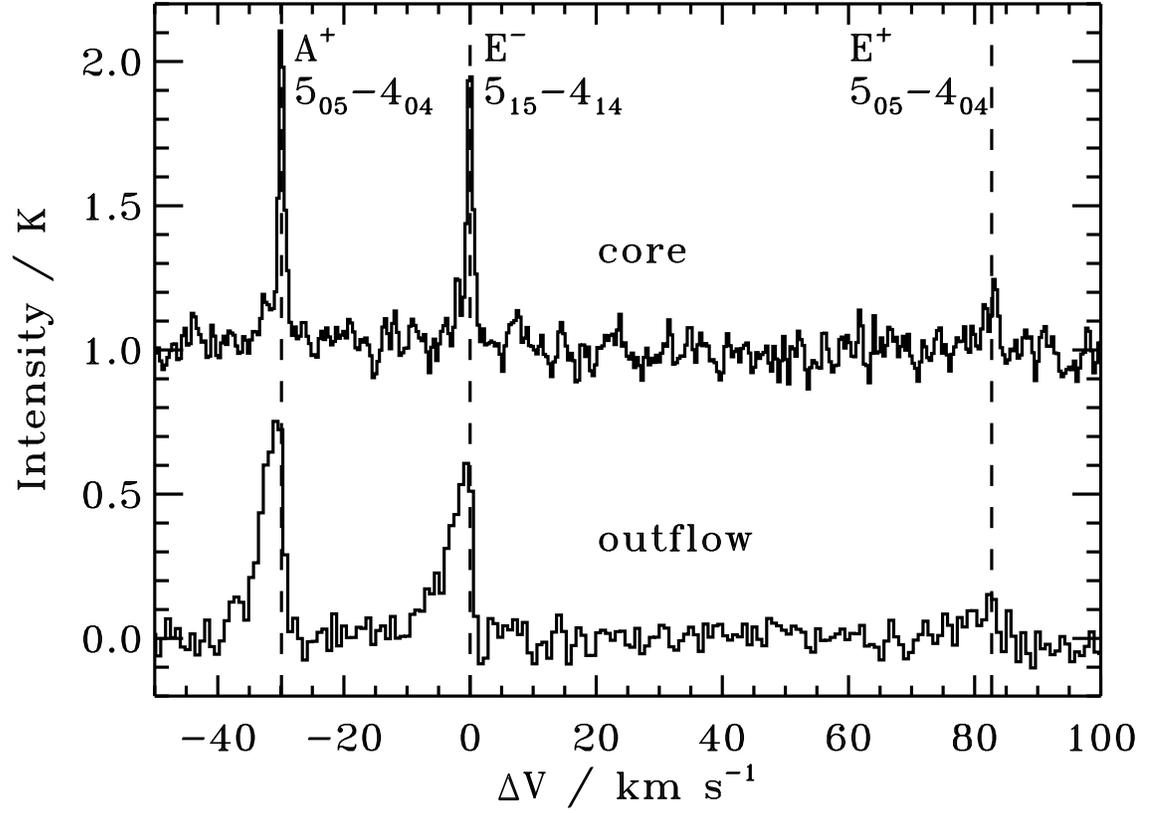}
\caption{Observed CH$_3$OH lines toward the `core' and `outflow' positions corrected by the `core' LSR velocity of 6.3 km s$^{-1}$ and plotted versus $\Delta$V, where 0 is the rest frequency 241.7672 GHz. \label{fig:ch3oh_sp}}
\end{figure}

Figure \ref{fig:ch3oh_sp} shows the extracted spectra at two of the CH$_3$OH intensity peaks in the channel map -- the `core' and southeast `outflow' positions. The CH$_3$OH emission toward the `core' consists of narrow, symmetric lines with small wings. Overall these spectra seem to be dominated by quiescent gas. In contrast the CH$_3$OH lines toward the 'outflow' position are more than four times broader and asymmetric, consistent with the assignment to an outflow. The spectra toward the `core' are similar to those observed by \citet{Hiramatsu10} toward the same line of sight. They did however not find any evidence for outflow activity toward B1-b. The lack of channel maps in \citet{Hiramatsu10} prevents direct comparison, but the detected outflows here are on a much smaller scale compared to the outflows they do observe and may thus have been overlooked.

Table \ref{tab:ch3oh} lists the observed CH$_3$OH transitions as well as the line FWHM and integrated intensities toward the core and outflow positions. To facilitate comparison with observations of complex molecules and previous CH$_3$OH observations at lower frequencies, the CH$_3$OH intensities were extracted both directly from the HERA maps and from maps convolved with 19'' and 27'' beams, the IRAM 30-m beam sizes at 131 and 89 GHz, respectively. Because of the asymmetry in the lines, the integrated intensities were calculated by integrating from -10 to +5 km s$^{-1}$ around each line center, rather than integrating  the fitted Gaussians used to estimate the FWHM. 

\begin{figure}[htp]
\plotone{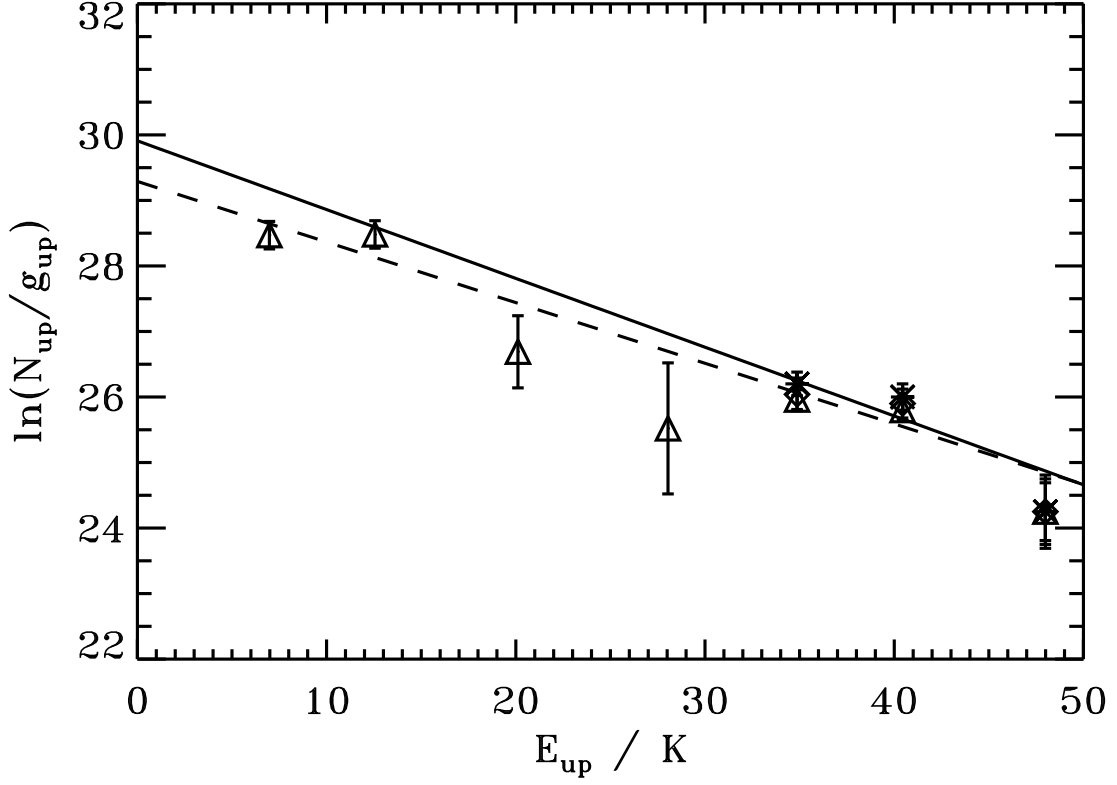}
\caption{Rotational diagrams of the CH$_3$OH abundances, derived from the map convolved with a 27'' beam, toward the `core' (diamonds), the `outflow' (stars) and the `protostar' (triangles), including previously published values \citep{Oberg09a}. The dashed line is fitted to the `protostar' abundances and the solid line to the 'core' abundances. \label{fig:ch3oh_rot}}
\end{figure}

The excitation temperature of CH$_3$OH toward each position is estimated using rotational diagrams \citep{Goldsmith99} in Fig. \ref{fig:ch3oh_rot} based on the CH$_3$OH map intensities convolved with a 27'' beam. The `core' and `outflow' positions only contain three CH$_3$OH lines each and the derived temperatures are therefore uncertain by at least a factor of two. The best fits results in 10~K for the `core' and 8~K for the `outflow'. For the `protostar' position the new data are combined with previous IRAM 30-m data observed at $\sim$97 GHz \citep{Oberg09a}. The resulting excitation temperature is $11\pm1$~K, confirming the estimated rotational temperatures toward the `core' and `outflow'. At the densities expected in molecular cloud cores, CH$_3$OH is easily sub-thermally excited -- rotational temperatures below 25~K are expected at densities up to 10$^6$ cm$^{-3}$ even if the kinetic temperature is above 100~K \citep{Bachiller98}. NH$_3$ maps of the same region however confirm that in this case the kinetic temperature is close to the CH$_3$OH rotational temperature, $\sim$10--12~K \citep{Bachiller90}.

\subsection{Search for complex molecules}

Building on the CH$_3$OH map, we searched for complex molecules toward the `core' and `outflow' positions. All acquired spectra are shown in the Appendices. Figure \ref{fig:comp_sp} shows six blow-ups focusing on the targeted complex molecules. The two lowest lying transitions  of HCOOCH$_3$ ($E_{\rm up}$ = 18~K) and of CH$_3$CHO ($E_{\rm up}$ = 28~K) are detected toward the core region (Fig. \ref{fig:comp_sp}a,d) at the 5-$\sigma$ level. Higher lying transitions of the two molecules are not detected at the 3-$\sigma$ level, nor are any transitions of C$_2$H$_5$OH, HOCH$_2$CHO or C$_2$H$_5$CN. One 2-$\sigma$ transition of CH$_3$OCH$_3$ is tentatively seen and its intensity is consistent with non-detections of the other lines. No complex molecule lines are detected toward the outflow (Fig. \ref{fig:comp_sp}). The two securely detected species, HCOOCH$_3$ and CH$_3$CHO, both display narrow line profiles between 0.5--0.9 km s$^{-1}$. The HCOOCH$_3$ lines show no evidence of asymmetry, while the  CH$_3$CHO lines have high-velocity wings suggestive of an emission contribution from the 'O2' outflow. 

All data for the targeted lines are listed in Table \ref{tab:comp}, including molecule transitions, catalogue frequencies, $E_{\rm low}$, FWHM of detected lines, integrated intensities for detected lines and integrated intensity upper limits for non-detections. The upper limits on higher lying HCOOCH$_3$ and CH$_3$CHO lines are consistent with the $\sim$10~K excitation temperature of CH$_3$OH toward the same line of sight. Thus a 10~K excitation temperature is assumed when calculating column densities.

Using these excitation temperatures, the total HCOOCH$_3$ column density (A + E) toward the  B1-b core is $8.3\pm2.8\times10^{12}$ cm$^{-2}$, averaged over the telescope beam, and the total (A + E) CH$_3$CHO column density is $5.4\pm1.4\times10^{12}$ cm$^{-2}$, which corresponds to abundances with respect to CH$_3$OH of 2.3 and 1.1\% respectively. The different beam sizes at different frequencies are accounted for by comparing the HCOOCH$_3$ to the CH$_3$OH column density derived with a 27'' beam and the CH$_3$CHO to the CH$_3$OH column density derived with the 19'' beam. The upper limit abundances for the other molecules are $\lesssim$0.8, $<$1.0\% and $<$1.1\% for CH$_3$OCH$_3$ , C$_2$H$_5$OH and HOCH$_2$CHO toward the core. The most significant upper limit toward the outflow is the total HCOOCH$_3$ abundance of $<$1.4\% with respect to CH$_3$OH. The calculated abundances and upper limits of complex molecules are summarized in Table \ref{tab:abund}. The lack of detections toward the outflow does not prove that it contains significantly less complex molecules than the `core', but the outflow position is definitely not richer in complex organics than the core region. The outflow upper limits are comparable to the detected abundances toward another low-mass outflow L1157 \citep{Arce08}

\section{Discussion}

\subsection{The origin of the CH$_3$OH gas}

The CH$_3$OH average abundance toward the quiescent CH$_3$OH `core' is $\sim2\times10^{-9}$ with respect to N$_{\rm H_2}$, assuming a SCUBA dust intensity conversion factor of N$_{\rm H_2}/S_{\rm \nu}^{\rm beam}$ of 1.3$\times10^{20}$ cm$^{-2}$ (mJy beam$^{-1}$)$^{-1}$ \citep{Kauffmann08}. This abundance is too high to explain with gas-phase formation of CH$_3$OH through any mechanism investigated so far, i.e. gas phase reactions in quiescent cores result in CH$_3$OH abundances below 10$^{-13}$ \citep{Garrod07}. The observed CH$_3$OH gas is thus a product of ice desorption.

The protostar B1-b belongs to a small set of young low-mass stellar objects with CH$_3$OH ice abundances of $>$10\% with respect to H$_2$O \citep[][Bottinelli et al. submitted to A\&A]{Boogert08}. CH$_3$OH ice has been previously observed to vary on small scales, however, and the ice abundance towards the `core' and `outflow' may be as low as 1\% with respect to H$_2$O ice. Still, $<$0.5\% of the CH$_3$OH ice must evaporate to account for the observed CH$_3$OH abundance of $\sim$2$\times10^{-9}$ toward the `core',  assuming a H$_2$O abundance of $\sim5\times10^{-5}$ with respect to H$_2$. 

The extent of the CH$_3$OH gas ($\sim$10000$\times$15000 AU) together with the low excitation temperature of CH$_3$OH and the offset between the CH$_3$OH `core' and the  `protostar' exclude thermal ice evaporation as a major contributor to the observed abundances -- a low-mass protostar cannot heat more than a few 100 AU to the sublimation temperature of CH$_3$OH of $\sim$80~K. This leaves non-thermal evaporation of ices, which can be broadly divided into two categories: ice evaporation due to grain sputtering in shocks and less violent non-thermal ice evaporation due to UV irradiation, cosmic rays or release of chemical energy \citep{Jones96, Shen04, Garrod07}. Grain sputtering is often invoked to explain excess CH$_3$OH in protostellar outflows \citep[e.g.][]{Bachiller01} and is probably responsible for the factor of a few higher abundances of CH$_3$OH gas in the two outflow positions around the dust core, using the same conversion factor as above to derive the dust column densities from the SCUBA map. The narrow CH$_3$OH lines in the remainder of the map suggest a different origin for the gas toward the protostar and `core' regions. 

The asymmetric distribution of CH$_3$OH compared to the dust core further suggest that ice is desorbed by irradiation from the direction of the protostar, illuminating the south-west side of the core. The exact nature of the irradiation is difficult to constrain without detailed modeling, but UV photodesorption is known to be efficient \citep{Oberg09d}. The dust core has a visual extinction of $>$100 mag. For radiation to affect the `core' region thus requires extensive cavities in the interstellar medium. As discussed in the Introduction, there are several large-scale outflows crossing the B1 region, which could have carved out such cavities \citep{Jorgensen06, Walawender09}. In particular Fig. 1 shows that there is a visible outflow in the 'right' direction originating either from the B1-b protostar or from the nearby B1-d protostar to the southwest.  In either case, the outflow may continue on towards the 'core', hidden from observations by the large column of dust, resulting in UV irradiation of the icy grains on the side of the dust core closest to the protostar. 

UV irradiation can be generated in a number or ways within such an outflow cavity. If an outflow is still active in the direction of the `core' position, the UV radiation may originate in the jet shock itself \citep{Reipurth01}. Some outflows are also known to become hot enough to emit x-rays \citep{Pravdo01}, which may directly desorb the ice or desorb it through secondary UV photons. The narrow line width of the CH$_3$OH emission suggest however that the CH$_3$OH gas does not originate close to shocked gas. It seems more likely that a past outflow has opened a cavity between the protostar and the `core' position and that UV radiation from this neighboring low-mass protostar protostar is desorbing the ices. Low-mass protostars are known to have excess UV fluxes compared to the interstellar radiation field, originating in the boundary layer between the protostar and the accretion disk \citep{Spaans95}.

UV photodesorption is thus a probable source of the CH$_3$OH gas distribution towards the B1-b dust core. Observations of UV tracers are however needed to confirm this scenario, since we cannot completely rule out that the CH$_3$OH are the left-overs from grain-sputtering by the same shock that opened up the cavity, if it passed through the area 10$^4$-10$^5$ years ago, the typical depletion time-scale at molecular cloud densities. 

\subsection{The origin of complex molecules toward B1-b}

The similar line widths and excitation temperature upper limits of HCOOCH$_3$ and CH$_3$CHO compared to CH$_3$OH suggest that the complex molecules toward the B1-b core originate from photodesorbed ice as well. This is a new potential source of complex organic molecules around low-mass protostars compared to what has been previously suggested in the literature, where complex molecules have been observed in a small warm core or disk where ices have evaporated thermally \citep[e.g.][]{Bottinelli07,Jorgensen05} or in shocks following grain sputtering \citep[e.g.][]{Arce08}. In difference to sputtering and thermal desorption, which quickly destroys the entire ice mantle in a small region, UV photodesorption can release a small fraction of the ice over large quiescent regions. This offers the opportunity to study ice chemistry {\it in situ} as it evolves around protostars, as previously suggested in \citet{Oberg09a, Oberg09d}. 

Non-thermal desorption of HCOOCH$_3$ and CH$_3$CHO is only a probable source of the observed gas if the molecules can be produced efficiently on grains, however. As discussed in the Introduction, such a production channel exists. UV processing of CO:CH$_3$OH-rich ices produces complex molecules at abundances that are consistent with the observed abundances of a few \% with respect to CH$_3$OH. 

Moreover this formation path explains the relative abundances of HCOOCH$_3$ and CH$_3$OCH$_3$ or C$_2$H$_5$OH. Figure \ref{fig:reaction} illustrates how the addition of CO affects the complex CH$_3$OH chemistry to produce the observed molecules. Quantitatively, UV irradiation of a CH$_3$OH:CO 1:10 ice mixture at 20~K under laboratory conditions results in the production of a CH$_3$OCH$_3$/HCOOCH$_3$ ratio of $<$0.1 and irradiation of a pure CH$_3$OH ice in CH$_3$OCH$_3$/HCOOCH$_3$ ratios of $>$1.3. While the laboratory results cannot be directly extrapolated to B1-b, a CO-rich ice is certainly needed to produce the observed low CH$_3$OCH$_3$(C$_2$H$_5$OH)/HCOOCH$_3$ ratios; without CO in the ice, the most recent model predict ratios of $>$10, rather than the observed $<$0.5 \citep{Garrod08}. 

\subsection{Comparison with previous complex chemistry observations}

With energetic CO:CH$_3$OH ice processing followed by non-thermal desorption established as the most probable explanation for the observed complex molecules in B1-b, it is interesting to see how the chemistry toward B1-b compares with complex chemistry observations toward other low-mass sources: NGC1333-IRAS 2A, 4A and 4B, IRAS 16293-2422 and L1157. All sources have been observed with single-dish telescopes where the emission region is assumed to be unresolved. These observations thus contain emission from both hot and cold material -- dependent on the density profile one or the other component may dominate. In addition IRAS 16293-2422 has been studied extensively using interferometry. In these observations two cores, A and B, are resolved and emission from the cold envelope can be excluded.

Because of assumptions on the emitting region for complex molecules in most previous studies, comparisons of absolute abundances are difficult between different sources using literature values.  Assuming the same source sizes of CH$_3$OH and complex molecules toward protostars observed with single-dish telescopes result in abundances of a few \% with respect to CH$_3$OH, comparable to those observed toward B1-b \citep{Maret05, Bottinelli04a, Bottinelli07}. Where beam averaged or resolved values of complex molecule abundances are actually reported, e.g. toward the low-mass outflow L1157 and the hot cores (or accretion disks) of the two IRAS 16293-2422 cores, the abundances are also of the order of a few \% \citep{Bisschop08, Arce08}. There is thus no evidence that the complex molecule abundances with respect to CH$_3$OH toward B1-b are special compared to other low-mass sources, ranging from outflows to hot cores. 

Table \ref{tab:compar} lists the abundance ratios of different complex molecules toward the different low-mass sources. The B1-b CH$_3$CHO/HCOOCH$_3$ ratio falls in between what has been previously measured toward the A and B cores in IRAS 16293. The upper limits on the CH$_3$OCH$_3$/ HCOOCH$_3$ and C$_2$H$_5$OH/HCOOCH$_3$ ratios are comparable to the single-dish observations toward NGC1333 2A, 4A and 4B, and possibly toward the IRAS 16293 envelope as well, but significantly lower compared to the resolved observations toward the IRAS 16293 A and B cores. In resolved observations toward NGC 1333 2A there is also a CH$_3$OCH$_3$ detection pointing to a higher CH$_3$OCH$_3$ to HCOOCH$_3$ ratio in the warm region close to the protostar \citep{Jorgensen05}. This is entirely consistent with a scenario where complex molecules form in cold CO:CH$_3$OH ice mixtures in the outer envelope, followed by further formation of CH$_3$OHCH$_3$ and C$_2$H$_5$OH in the warmer CO-depleted CH$_3$OH ice close to the protostar (Fig. \ref{fig:cartoon}). In the cold parts complex ices are only released non-thermally through e.g. UV photodesorption, while closer to the protostar thermal desorption is likely to dominate. This agrees with the low excitation temperatures in the single dish observations compared to the observations that resolve the innermost part of the envelope. 

\section{Conclusions}

\begin{enumerate}
\item Cold CH$_3$OH gas (excitation temperature of $\sim$10~K) is abundant and widespread toward B1-b. Quiescent CH$_3$OH is most abundant in between the B1-b SCUBA dust core and the B1-b protostar, indicative of UV photodesorption of ice on the side of the quiescent dust core because of radiation from the protostar escaping through outflow cavities.
\item Cold HCOOCH$_3$ and CH$_3$CHO are both detected and CH$_3$OCH$_3$ is tentatively detected toward the quiescent CH$_3$OH peak. No complex molecules are observed toward an outflow associated with B1-b, but 3-$\sigma$ upper limits with respect to CH$_3$OH are only slightly lower compared to the quiescent core. In addition an asymmetry in the CH$_3$CHO emission lines is suggestive of some contribution from a small outflow included in the beam.
\item Assuming a 10~K excitation temperature the calculated beam-averaged abundances are 2.3\% for HCOOCH$_3$, 1.1\% for CH$_3$CHO, $\lesssim$0.8\% for CH$_3$OCH$_3$ and $<$1.0-1.1\% for C$_2$H$_5$OH and HOCH$_2$CHO with respect to CH$_3$OH. 
\item Building on recent experiments, the observations of large abundances of HCO-containing molecules are explained by UV/cosmic ray processing of cold CH$_3$OH:CO ice followed by non-thermal desorption of a fraction of the produced organic ice.
\item The beam-averaged abundances with respect to CH$_3$OH and the ratios between different complex molecules are similar to other single-dish observations toward low-mass protostars. In contrast resolved observations of the cores around protostars are more abundandant in complex molecules without a HCO-group. This is consistent with a complex ice chemistry that evolves from HCO-rich to HCO-poor as the ice warms up and the CO-ice evaporates when the icy dust falls toward  the protostar.
\end{enumerate}

Including B1-b, there are still only a handful of observations of complex molecules toward low-mass protostars. In addition few complex molecules are detected toward each source. This makes it difficult both to determine the prevalence of a complex organic chemistry around low-mass protostars and to firmly establish the main formation path of such molecules. The detections of complex molecules in the vicinity of the CH$_3$OH-ice-rich protostar suggests that targeting other CH$_3$OH-ice rich protostars may increase our sample of complex molecule detections significantly. Maybe more important from a formation pathway point of view is to increase our understanding on how the complex chemistry varies between warm, luke-warm and cold regions i.e. how it varies with distance from the protostar. This should be pursued both on larger scales with single dish observations and on smaller scales with interferometers -- similarly to what has been done toward IRAS 16293 and NGC1333 2a \citep[e.g.][]{Jorgensen05,Bisschop08}. In the meantime the complex molecules observed toward B1-b, and their exceptionally low excitation temperatures and line widths, provides clear evidence, for the efficient formation of complex ices during star- and planet-formation.

\acknowledgments

We are grateful to the IRAM staff for help with the observations and reduction of the resulting data. This work has benefitted from discussions with Herma Cuppen, Robin Garrod and Lars Kristensen and from comments by an anonymous referee. Umut Yildiz carried out the second round of observations at the IRAM 30-m with the financial support of RadioNet. Support for K.~I.~\"O is provided by NASA through Hubble Fellowship grant  awarded by the Space Telescope Science Institute, which is operated by the Association
of Universities for Research in Astronomy, Inc., for NASA, under contract NAS 5-26555. Astrochemistry in Leiden is supported by a SPINOZA grant of the Netherlands Organization for Scientific Research (NWO). The Centre for Star and Planet Formation is funded by the Danish National Research Foundation and the University of Copenhagen's programme of excellence.

\begin{figure}[htp]
\epsscale{1.0}
\plotone{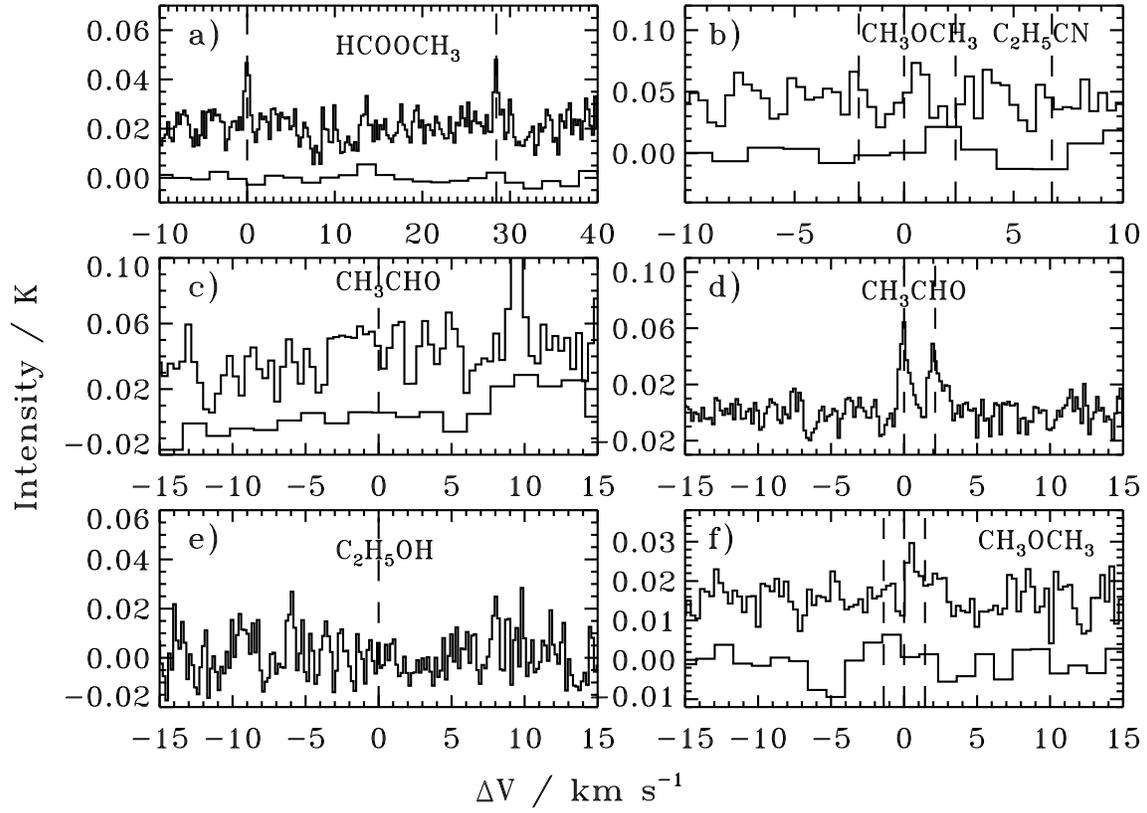}
\caption{Detected HCOOCH$_3$ and CH$_3$CHO lines, a tentative CH$_3$OCH$_3$ detection in panel f) and significant non-detections toward the `core' position (upper spectra) and the `outflow' position when observed (lower spectra). The spectra are centered on the rest frequency of a) HCOOCH$_3$-E $7_{1\:6}-7_{1\:5}$, b) CH$_3$OCH$_3$ $5_{1\:5\:1}-5_{0\:4\:1}$, c) CH$_3$CHO-E $6_{2\:5}-5_{2\:4}$, d) CH$_3$CHO-E $7_{17}-6_{16}$, e) C$_2$H$_5$OH $4_{3\:2\:2}-4_{2\:3\:2}$ and f) CH$_3$OCH$_3$ $7_{1\:7\:2}-6_{0\:6\:2}$ assuming a systematic velocity of 6.3 km s$^{-1}$. The unmarked line in panel c is real and due to NS (see Appendices) .\label{fig:comp_sp}}
\end{figure}

\begin{figure}[htp]
\epsscale{1.0}
\plotone{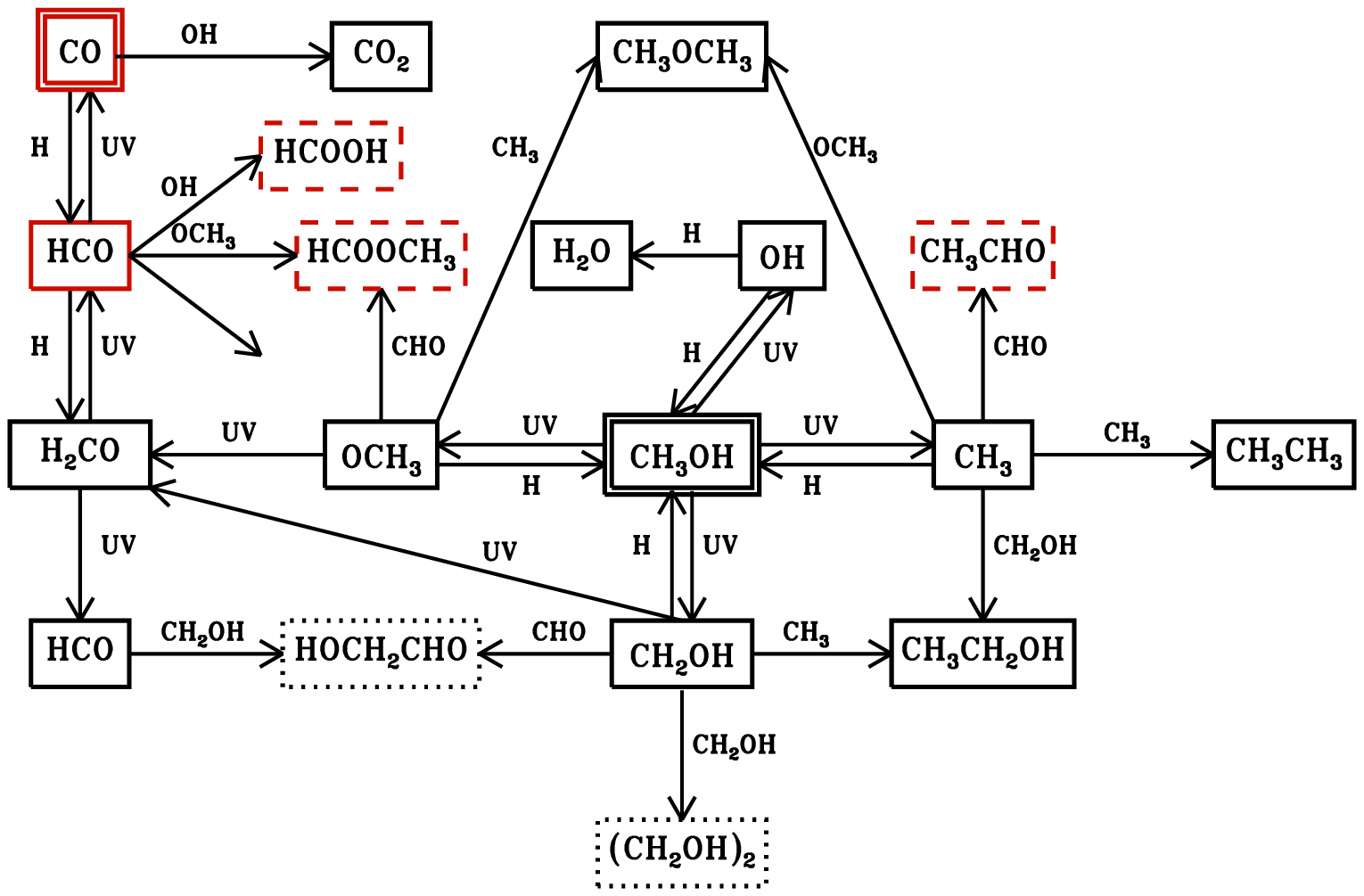}
\caption{Reaction scheme of a CH$_3$OH-based ice photochemistry adapted from \citet{Oberg09d} to high-light the impact of CO. In a CO-dominated ice the production of several HCO-bearing complex species (red dashed boxes) is enhanced relative to molecules such as C$_2$H$_5$OH, CH$_3$OCH$_3$ and (CH$_2$OH)$_2$, which tend to dominate among the pure CH$_3$OH-photochemistry products (solid and dotted boxes). (CH$_2$OH)$_2$ and HOCH$_2$CHO (dotted boxes) are only produced abundantly at luke-warm temperatures ($>$30~K in the laboratory). \label{fig:reaction}}
\end{figure}
 
\begin{figure}[htp]
\epsscale{1.0}
\plotone{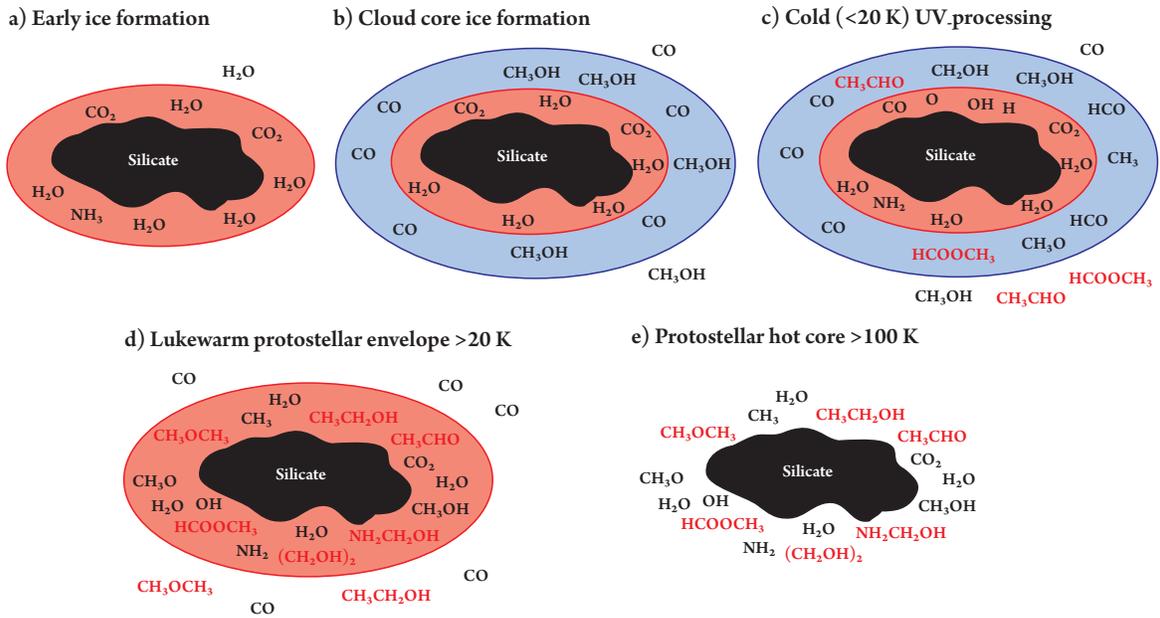}
\vspace{1cm}
\caption{The suggested evolution of ices during star formation. Pink indicates a H$_2$O-dominated ice and blue a CO-dominated ice. At each cold stage a small amount of the ice is released non-thermally. Early during cloud formation (a) a H$_2$O-rich ice forms. Once a critical density and temperature is reached CO freezes out catastrophically (b), providing reactants for CH$_3$OH ice formation. Far away from the protostar (c), photoprocessing of the CO-rich ice results in the production of e.g. HCOOCH$_3$. Closer to the protostar (d), following sublimation of CO, other complex molecules become abundant. Finally all ice desorb thermally close to the protostar $>$100~K (e). \label{fig:cartoon}}
\end{figure}

\newpage

\begin{table}[htp]
\vspace{1cm}
{\footnotesize
\begin{center}
\caption{Observational parameters\label{tab:obs}}
\bigskip

\begin{tabular}{lcccc}
\hline\hline
\multicolumn{5}{c}{Positions and CH$_3$OH observations with HERA}\\
\hline
Name\tablenotemark{a} & R.A. (J2000) & Dec. (J2000) & rms (mK) & $\delta V$ (km s$^{-1}$)\\
\hline
`core' & 03:33:20.80&31:07:40.0&  92 &0.39\\
`outflow' & 03:33:21.90& 31:07:22.0 & 65 &0.77\\
`protostar' &03:33:20.34 & 31:07:21.4 & 92 &0.39\\
\hline
\end{tabular}

\bigskip

\begin{tabular}{ccccc c cc}
\hline \hline
\multicolumn{8}{c}{Complex molecules observations with EMIR} \\
\hline
Setting & Receiver & Frequency band (GHz) & \multicolumn{2}{c}{rms (mK)} && \multicolumn{2}{c}{$\delta V$ (km s$^{-1}$)} \\
\cline{4-5}\cline{7-8}
& & & `core' & `outflow' & & `core' & `outflow' \\
\hline
1 & E90 & 88.81--88.93 & 5.4 & 2.0 && 0.26 & 2.1 \\
  & E150 & 146.84--146.96 & 5.5 & 2.2 && 0.32 & 1.3 \\
2 & E90 & 115.46--115.58 & 14.6 & 9.9 && 0.41 & 1.6 \\
  & E150 & 146.93--147.05 & 4.6 & 3.9 && 0.32 & 1.3 \\
3 & E150 & 130.82--130.94 & 8.4 & ... && 0.18 & ... \\
\hline
\end{tabular}
\tablenotetext{a}{The `core' is the quiescent CH$_3$OH maximum $\sim$10 offset from the SCUBA core and the `protostar' is the source discovered by {\it Spitzer}.}
\end{center}
}
\end{table}

\begin{table}[ht]
\begin{center}
{\footnotesize
\caption{Observed CH$_3$OH lines with 1-$\sigma$ uncertainties in brackets. \label{tab:ch3oh}}
\begin{tabular}{lccccccc}
\hline \hline
Position &Transition & Freqency & $E_{\rm low}$ & FWHM  & $\int T dV$ 10''  & $\int T dV$ 19''  & $\int T dV$ 27'' \\
& & GHz & cm$^{-1}$ & km s$^{-1}$ & K km s$^{-1}$& K km s$^{-1}$& K km s$^{-1}$\\
\hline`core' 	&A$^+5_{05}-4_{04}$ &241.7914&16.1 & 1.2[0.1] 	& 2.8[0.6]&1.5[0.3]&1.2[0.3]\\
& E$^-5_{15}-4_{14}$ &241.7672&20.0 &  1.3[0.1] 	& 2.2[0.4]&1.2[0.2]&1.1[0.2]\\
\smallskip
&E$^-5_{05}-4_{04}$ &241.7002&25.3 &  2.4[0.5] 		& 0.5[0.1]&0.2[0.1]&0.2[0.1]\\
`outflow' 	&A$^+5_{05}-4_{04}$ &241.7914&16.1 & 4.0[0.2] 		& 3.3[0.7]&1.9[0.4]&1.4[0.4]\\
& E$^-5_{15}-4_{14}$ &241.7672&20.0 &  4.4[0.3] 	& 2.8[0.5]&1.6[0.3]&1.2[0.3]\\
\smallskip
&E$^-5_{05}-4_{04}$ &241.7002&25.3 &  5.7[1.6] 		& 0.5[0.1]&0.2[0.1]&0.2[0.1]\\
`protostar' &A$^+5_{05}-4_{04}$ &241.7914&16.1 & 1.1[0.1] & 1.6[0.3]&1.2[0.3]&1.2[0.3]\\
& E$^-5_{15}-4_{14}$ &241.7672&20.0 & 1.0[0.1] 	& 1.4[0.3]&1.1[0.2]&1.0[0.2]\\
&E$^-5_{05}-4_{04}$ &241.7002&25.3 & 1.0[0.9] 		& 0.5[0.1]&0.2[0.1]&0.2[0.1]\\
\hline
\end{tabular}
}
\end{center}
\end{table}

\begin{table}[ht]
\begin{center}
{\footnotesize
\caption{Observed complex molecular lines with 1-$\sigma$ uncertainties in brackets. \label{tab:comp}}
\begin{tabular}{lccccc}
\hline \hline
Molecule & Transition & Freq. & $E_{\rm low}$   &FWHM\tablenotemark{a} & $\int T dV$\tablenotemark{b}\\
 &   & GHz&  cm$^{-1}$  & km s$^{-1}$ &  mK km s$^{-1}$\\ 
 \hline
HCOOCH$_3$-E & $7_{1\:6}-7_{1\:5}$ &88.8432 &9.5  &0.55[0.13]  & 18[4] / $<24$\\
HCOOCH$_3$-A & $7_{1\:6}-7_{1\:5}$ &88.8948 &9.5 &0.60[0.12]  & 16[4] / $<24$\\
C$_2$H$_5$CN &$19_{1\:18}-19_{0\:19}$& 88.8948 &55.6 & \nodata  &$<11$ / $<24$\\
HOCH$_2$CHO &$9_{4\:6}-9_{s\:7}$ &88.8924 &21.2  & \nodata  & $<11$ / $<24$\\

CH$_3$CHO-A &$6_{2\:5}-5_{2\:4}$ &115.4939 &15.9  & \nodata& $<30$ / $<119$\\
C$_2$H$_5$OH &$5_{1\:5}-5{0\:5}$ &115.4945 &48.2  & \nodata &$<30$ / $<119$\\
CH$_3$OCH$_3$ &$5_{1\:5\:2}-5_{0\:4\:2}$ &115.5440 &6.3  & \nodata& $<30$ / $<119$\\
CH$_3$OCH$_3$ &$5_{1\:5\:3}-5_{0\:4\:3}$ &115.5440 &6.3 & \nodata& $<30$ / $<119$\\
CH$_3$OCH$_3$ &$5_{1\:5\:1}-5_{0\:4\:1}$ &115.5448 &6.3  & \nodata& $<30$ / $<119$\\
CH$_3$OCH$_3$ &$5_{1\:5\:0}-5_{0\:4\:0}$ &115.5457 &6.3  & \nodata& $<30$ / $<119$\\
C$_2$H$_5$CN &$6_{3\:3}-6_{2\:4}$&115.5474 &9.4 & \nodata& $<30$ / $<119$\\

C$_2$H$_5$OH &$4_{3\:2\:2}-4_{2\:3\:2}$ &130.8715 &9.3 &  \nodata & $<17$\\
CH$_3$CHO-E &$7_{1\:7}-6_{1\:6}$ &130.8918 &14.7 &0.89[0.18] &39[6]\\
CH$_3$CHO-A &$7_{1\:7}-6_{1\:6}$ &130.8927 &14.7 &0.67[0.18] &42[6]\\
C$_2$H$_5$CN &$15_{0\:15}-14_{0\:14}$ &130.9039 &30.9 &  \nodata & $<17$\\

CH$_3$OCH$_3$ &$5_{3\:3\:2}-5_{2\:4\:2}$ &146.8660 &13.4  & \nodata& $<11$ / $<26$\\
CH$_3$OCH$_3$ &$5_{3\:3\:3}-5_{2\:4\:3}$ &146.8703 &13.4  & \nodata& $<11$ / $<26$\\
CH$_3$OCH$_3$ &$5_{3\:3\:1}-5_{2\:4\:1}$ &146.8725 &13.4 & \nodata& $<11$ / $<26$\\
CH$_3$OCH$_3$ &$5_{3\:3\:0}-5_{2\:4\:0}$ &146.8773 &13.4 & \nodata& $<11$ / $<26$\\
NH$_2$CHO &$7_{0\:7\:8}-6_{0\:6\:7}$ &146.8716 &14.8  & \nodata& $<11$ / $<26$\\
NH$_2$CHO &$7_{0\:7\:6}-6_{0\:6\:5}$ &146.8716 &14.8  & \nodata& $<11$ / $<26$\\
NH$_2$CHO &$7_{0\:7\:7}-6_{0\:6\:6}$ &146.8717 &14.8  & \nodata& $<11$ / $<26$\\
C$_2$H$_5$CN &$17_{1\:17}-16_{1\:16}$ &146.8826 &40.1 & \nodata& $<11$ / $<26$\\

HCOOCH$_3$-E &$12_{3\:10}-11_{3\:9}$ &146.9777 &31.3  & \nodata&$<9$ / $<47$\\
HCOOCH$_3$-A &$12_{3\:10}-11_{3\:9}$ &146.9880 &31.2  & \nodata&$<9$ / $<47$\\
CH$_3$OCH$_3$ &$7_{1\:7\:2}-6_{0\:6\:2}$ &147.0242 &13.2  & \nodata&$<9$ / $<47$\\
CH$_3$OCH$_3$ &$7_{1\:7\:3}-6_{0\:6\:3}$ &147.0242 &13.2  & \nodata&$<9$ / $<47$\\
CH$_3$OCH$_3$ &$7_{1\:7\:1}-6_{0\:6\:1}$ &147.0249 &13.2  & $\sim$0.53 &$\sim$8 / $<47$\\
CH$_3$OCH$_3$ &$7_{1\:7\:0}-6_{0\:6\:0}$ &147.0256 &13.2 & \nodata&$<9$ / $<47$\\
\hline
\end{tabular}
\tablenotetext{b}{FWHM for detected lines toward the 'core' position.}
\tablenotetext{b}{Values are for `core' / `outflow' positions. 3-$\sigma$ upper limits are calculated from the rms assuming line widths of 0.68 (average of the four complex line detections) and 4.2 km s$^{-1}$ for the `core' and `outflow' positions, respectively.}
}
\end{center}
\end{table}

\begin{table}[ht]
\begin{center}
{\footnotesize
\caption{Calculated abundances and 3-$\sigma$ upper limits with 1-$\sigma$ uncertainties in brackets.\label{tab:abund}}
\begin{tabular}{lcccc}
\hline \hline
Position & Species & T$_{\rm rot}$ / K &  N\tablenotemark{a} / 10$^{13}$ cm$^{-2}$ &\% CH$_3$OH\tablenotemark{a}\\
\hline
`core' 	&CH$_3$OH &10[5] &$\sim$36 / 47\tablenotemark{b} &100\\
`core' 	&HCOOCH$_3$-E & -- &0.44[0.10] &1.2[0.3]\\
`core' 	&HCOOCH$_3$-A & -- &0.39[0.10] &1.1[0.3]\\
`core'		&CH$_3$CHO-E     & -- &0.26[0.04] &0.6[0.1]\\
`core'		&CH$_3$CHO-A     & -- &0.28[0.04] &0.6[0.1]\\
`core'		&CH$_3$OCH$_3$&--&$\lesssim$0.36 &$\lesssim$0.8\\
`core'		&C$_2$H$_5$OH&--&$<$0.48 &$<$1.0\\
`core'		&HCOCH$_2$OH&--&$<$0.41 &$<$1.1\\
`outflow' 	&CH$_3$OH &8[4] &$\sim$78 / 83\tablenotemark{b} &100\\
`outflow' 	&HCOOCH$_3$-E &--&$<$0.56 &$<$0.7\\
`outflow' 	&HCOOCH$_3$-A &--&$<$0.56 &$<$0.7\\
`protostar' &CH$_3$OH &11[1] &23[4] &100\\
\hline
\end{tabular}
\tablenotetext{a}{Assuming an excitation temperature of 10~K for the complex species toward both the core and the outflow position.}
\tablenotetext{b}{The first value is derived assuming a 27'' beam and the second assuming a 19'' beam.}
}
\end{center}
\end{table}

\begin{table}[ht]
\begin{center}
{\footnotesize
\caption{Calculated abundances and 3-$\sigma$ upper limits with 1-$\sigma$ uncertainties in brackets.\label{tab:abund}}
\begin{tabular}{lccccc}
\hline \hline
Source & T$_{\rm rot}$ &  FWHM  &CH$_3$CHO&CH$_3$OCH$_3$ &C$_2$H$_5$OH\\
& K &  km s$^{-1}$ & / HCOOCH$_3$\tablenotemark{a} & / HCOOCH$_3$\tablenotemark{a} & / HCOOCH$_3$\tablenotemark{a}\\
\hline
B1-b core &10[10] & 0.55--0.89 & 0.5[0.3] &$\lesssim$0.3 &$<$0.4\\
IRAS 16293 env.\tablenotemark{b} &14--67 &0.8--10 & 0.1& 0.67 &--\\
IRAS 16293 core A / B \tablenotemark{c} &100--300 &1.5--2.5 & $<$0.1 / 3&1 / 2  &2 /  1\\
NGC1333 IRAS 2A\tablenotemark{d,e} &38--101 &1--4  &-- &$<$0.3&--\\
NGC1333 IRAS 4A\tablenotemark{d} &24--36 &1.2--4.2 &-- &$<$0.4 &--\\
NGC1333 IRAS 4B\tablenotemark{d} &34--38 &0.5--1.8 &-- &$<$0.7 &--\\
L1157 outflow\tablenotemark{f} &18--27 & 2.3--5.5 &-- & -- &0.4\\
\hline
\end{tabular}
\tablenotetext{a}{Where only A or E column densities are available the total HCOOCH$_3$ abundance is dervied by assuming equal abundances of both.}
\tablenotetext{b}{\citet{Cazaux03, Herbst09}}
\tablenotetext{c}{Beamsize of 5.5''$\times$3.2'' \citet{Bisschop08}, 1.3''$\times$2.7'' \citet{Kuan04} and \citet{Huang05}}
\tablenotetext{d}{Abundances are rescaled assuming the same beam dilution for CH$_3$OH and complex organic molecules. Single dish column densities from \citet{Maret05, Bottinelli04a, Bottinelli07} }
\tablenotetext{e}{CH$_3$OCH$_3$ is detected toward the central core using the Sub-Millimeter Array \citep{Jorgensen05}. }
\tablenotetext{f}{\citet{Arce08}}
}
\end{center}
\end{table}

\newpage


\begin{figure}[htp]
\epsscale{.75}
\plotone{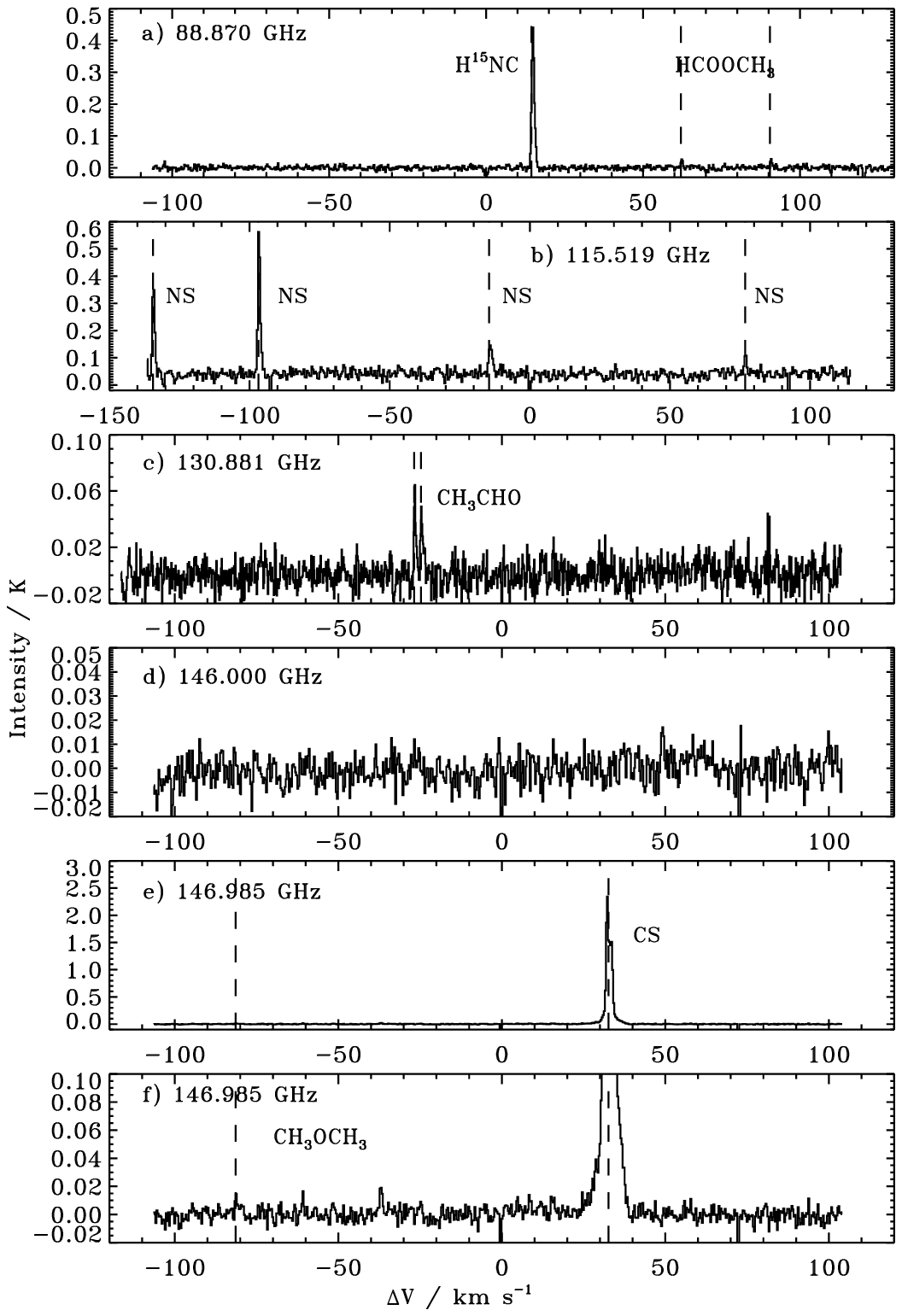}
\caption{Observed lines in the five different settings explored toward the 'core' position plotted versus $\Delta$V, the deviation from the core velocity 6.3 km s$^{-1}$. The spectra are centered on the rest frequency shown in each panel. \label{fig:app_sp_1}}
\end{figure}

\begin{figure}[htp]
\epsscale{.75}
\plotone{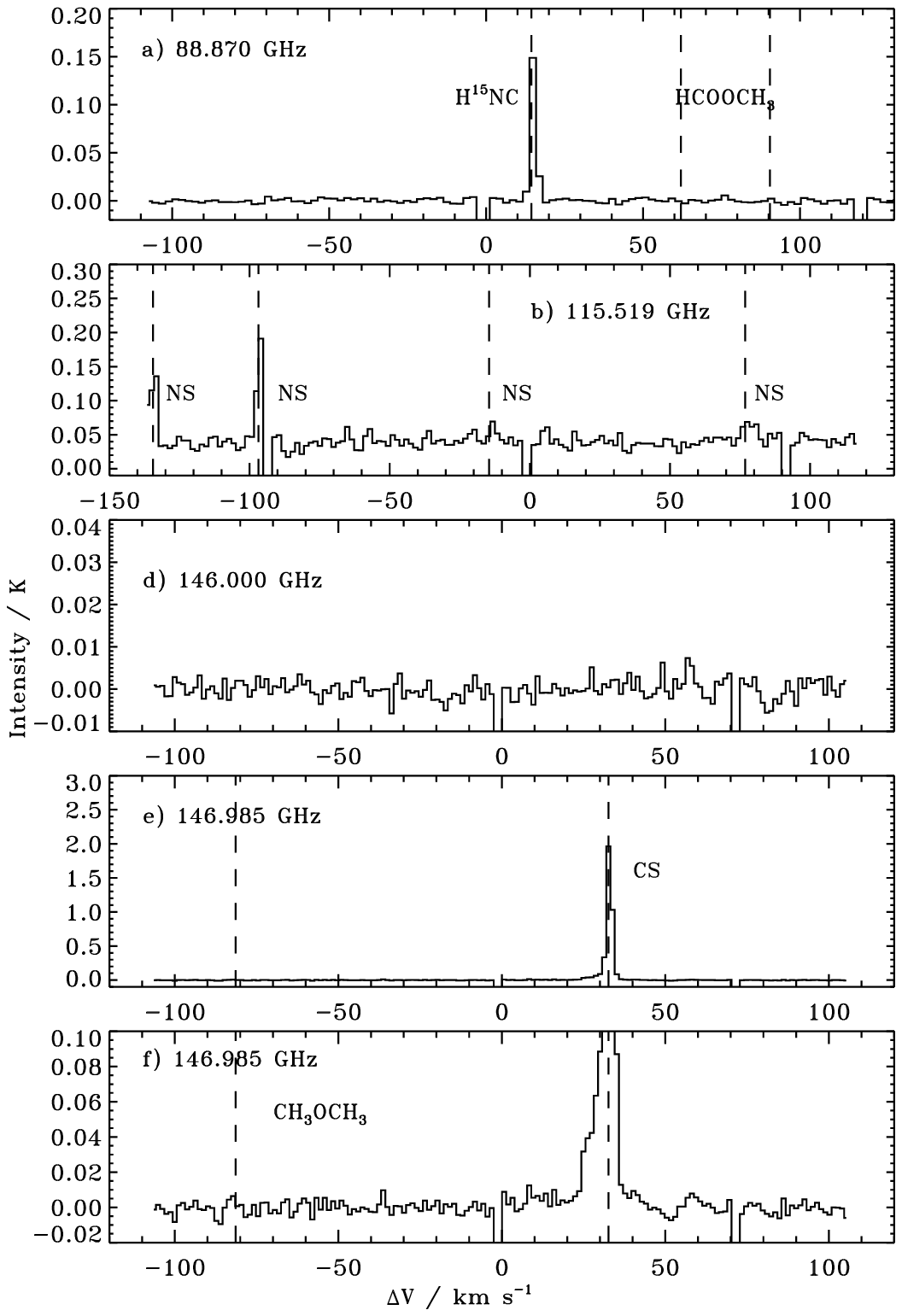}
\caption{Observed lines in the five different settings explored toward the 'outflow' position plotted versus $\Delta$V, the deviation from the core velocity 6.3 km s$^{-1}$. The spectra are centered on the rest frequency shown in each panel. \label{fig:app_sp_2}}
\end{figure}


\begin{thebibliography}{39}

\bibitem[{{Arce} {et~al.}(2008){Arce}, {Santiago-Garc{\'{\i}}a},
  {J{\o}rgensen}, {Tafalla}, \& {Bachiller}}]{Arce08}
{Arce}, H.~G., {Santiago-Garc{\'{\i}}a}, J., {J{\o}rgensen}, J.~K., {Tafalla},
  M., \& {Bachiller}, R. 2008, \apjl, 681, L21

\bibitem[{{Bachiller} {et~al.}(1998){Bachiller}, {Codella}, {Colomer},
  {Liechti}, \& {Walmsley}}]{Bachiller98}
{Bachiller}, R., {Codella}, C., {Colomer}, F., {Liechti}, S., \& {Walmsley},
  C.~M. 1998, \aap, 335, 266

\bibitem[{{Bachiller} {et~al.}(1990){Bachiller}, {Menten}, \& {del Rio
  Alvarez}}]{Bachiller90}
{Bachiller}, R., {Menten}, K.~M., \& {del Rio Alvarez}, S. 1990, \aap, 236, 461

\bibitem[{{Bachiller} {et~al.}(2001){Bachiller}, {P{\'e}rez Guti{\'e}rrez},
  {Kumar}, \& {Tafalla}}]{Bachiller01}
{Bachiller}, R., {P{\'e}rez Guti{\'e}rrez}, M., {Kumar}, M.~S.~N., \&
  {Tafalla}, M. 2001, \aap, 372, 899

\bibitem[{{Bennett} \& {Kaiser}(2007)}]{Bennett07b}
{Bennett}, C.~J. \& {Kaiser}, R.~I. 2007, \apj, 661, 899

\bibitem[{{Bisschop} {et~al.}(2006){Bisschop}, {Fraser}, {{\"O}berg}, {van
  Dishoeck}, \& {Schlemmer}}]{Bisschop06}
{Bisschop}, S.~E., {Fraser}, H.~J., {{\"O}berg}, K.~I., {van Dishoeck}, E.~F.,
  \& {Schlemmer}, S. 2006, \aap, 449, 1297

\bibitem[{{Bisschop} {et~al.}(2008){Bisschop}, {J{\o}rgensen}, {Bourke},
  {Bottinelli}, \& {van Dishoeck}}]{Bisschop08}
{Bisschop}, S.~E., {J{\o}rgensen}, J.~K., {Bourke}, T.~L., {Bottinelli}, S., \&
  {van Dishoeck}, E.~F. 2008, \aap, 488, 959

\bibitem[{{Boogert} {et~al.}(2008){Boogert}, {Pontoppidan}, {Knez}, {Lahuis},
  {Kessler-Silacci}, {van Dishoeck}, {Blake}, {Augereau}, {Bisschop},
  {Bottinelli}, {Brooke}, {Brown}, {Crapsi}, {Evans}, {Fraser}, {Geers},
  {Huard}, {J{\o}rgensen}, {{\"O}berg}, {Allen}, {Harvey}, {Koerner}, {Mundy},
  {Padgett}, {Sargent}, \& {Stapelfeldt}}]{Boogert08}
{Boogert}, A.~C.~A., {Pontoppidan}, K.~M., {Knez}, C., {et~al.} 2008, \apj,
  678, 985

\bibitem[{{Bottinelli} {et~al.}(2004){Bottinelli}, {Ceccarelli}, {Lefloch},
  {Williams}, {Castets}, {Caux}, {Cazaux}, {Maret}, {Parise}, \&
  {Tielens}}]{Bottinelli04a}
{Bottinelli}, S., {Ceccarelli}, C., {Lefloch}, B., {et~al.} 2004, \apj, 615,
  354

\bibitem[{{Bottinelli} {et~al.}(2007){Bottinelli}, {Ceccarelli}, {Williams}, \&
  {Lefloch}}]{Bottinelli07}
{Bottinelli}, S., {Ceccarelli}, C., {Williams}, J.~P., \& {Lefloch}, B. 2007,
  \aap, 463, 601

\bibitem[{{Cazaux} {et~al.}(2003){Cazaux}, {Tielens}, {Ceccarelli}, {Castets},
  {Wakelam}, {Caux}, {Parise}, \& {Teyssier}}]{Cazaux03}
{Cazaux}, S., {Tielens}, A.~G.~G.~M., {Ceccarelli}, C., {et~al.} 2003, \apjl,
  593, L51

\bibitem[{{Charnley} {et~al.}(1992){Charnley}, {Tielens}, \&
  {Millar}}]{Charnley92}
{Charnley}, S.~B., {Tielens}, A.~G.~G.~M., \& {Millar}, T.~J. 1992, \apjl, 399,
  L71

\bibitem[{{Cuppen} {et~al.}(2009){Cuppen}, {van Dishoeck}, {Herbst}, \&
  {Tielens}}]{Cuppen09}
{Cuppen}, H.~M., {van Dishoeck}, E.~F., {Herbst}, E., \& {Tielens}, A.~G.~G.~M.
  2009, \aap, 508, 275

\bibitem[{{Garrod} \& {Herbst}(2006)}]{Garrod06}
{Garrod}, R.~T. \& {Herbst}, E. 2006, \aap, 457, 927

\bibitem[{{Garrod} {et~al.}(2007){Garrod}, {Wakelam}, \& {Herbst}}]{Garrod07}
{Garrod}, R.~T., {Wakelam}, V., \& {Herbst}, E. 2007, \aap, 467, 1103

\bibitem[{{Garrod} {et~al.}(2008){Garrod}, {Weaver}, \& {Herbst}}]{Garrod08}
{Garrod}, R.~T., {Weaver}, S.~L.~W., \& {Herbst}, E. 2008, \apj, 682, 283

\bibitem[{{Goldsmith} \& {Langer}(1999)}]{Goldsmith99}
{Goldsmith}, P.~F. \& {Langer}, W.~D. 1999, \apj, 517, 209

\bibitem[{{Herbst} \& {van Dishoeck}(2009)}]{Herbst09}
{Herbst}, E. \& {van Dishoeck}, E.~F. 2009, \araa, 47, 427

\bibitem[{{Hiramatsu} {et~al.}(2010){Hiramatsu}, {Hirano}, \&
  {Takakuwa}}]{Hiramatsu10}
{Hiramatsu}, M., {Hirano}, N., \& {Takakuwa}, S. 2010, ArXiv e-prints

\bibitem[{{Hirano} {et~al.}(1999){Hirano}, {Kamazaki}, {Mikami}, {Ohashi}, \&
  {Umemoto}}]{Hirano99}
{Hirano}, N., {Kamazaki}, T., {Mikami}, H., {Ohashi}, N., \& {Umemoto}, T.
  1999, in Proceedings of Star Formation 1999, held in Nagoya, Japan, ed.
  {T.~Nakamoto}, 181--182

\bibitem[{{Huang} {et~al.}(2005){Huang}, {Kuan}, {Charnley}, {Hirano},
  {Takakuwa}, \& {Bourke}}]{Huang05}
{Huang}, H.-C., {Kuan}, Y.-J., {Charnley}, S.~B., {et~al.} 2005, Advances in
  Space Research, 36, 146

\bibitem[{{Jones} {et~al.}(1996){Jones}, {Tielens}, \& {Hollenbach}}]{Jones96}
{Jones}, A.~P., {Tielens}, A.~G.~G.~M., \& {Hollenbach}, D.~J. 1996, \apj, 469,
  740

\bibitem[{{J{\o}rgensen} {et~al.}(2005){J{\o}rgensen}, {Bourke}, {Myers},
  {Sch{\"o}ier}, {van Dishoeck}, \& {Wilner}}]{Jorgensen05}
{J{\o}rgensen}, J.~K., {Bourke}, T.~L., {Myers}, P.~C., {et~al.} 2005, \apj,
  632, 973

\bibitem[{{J{\o}rgensen} {et~al.}(2006){J{\o}rgensen}, {Harvey}, {Evans},
  {Huard}, {Allen}, {Porras}, {Blake}, {Bourke}, {Chapman}, {Cieza}, {Koerner},
  {Lai}, {Mundy}, {Myers}, {Padgett}, {Rebull}, {Sargent}, {Spiesman},
  {Stapelfeldt}, {van Dishoeck}, {Wahhaj}, \& {Young}}]{Jorgensen06}
{J{\o}rgensen}, J.~K., {Harvey}, P.~M., {Evans}, II, N.~J., {et~al.} 2006,
  \apj, 645, 1246

\bibitem[{{Kauffmann} {et~al.}(2008){Kauffmann}, {Bertoldi}, {Bourke}, {Evans},
  \& {Lee}}]{Kauffmann08}
{Kauffmann}, J., {Bertoldi}, F., {Bourke}, T.~L., {Evans}, II, N.~J., \& {Lee},
  C.~W. 2008, \aap, 487, 993

\bibitem[{{Kuan} {et~al.}(2004){Kuan}, {Huang}, {Charnley}, {Hirano},
  {Takakuwa}, {Wilner}, {Liu}, {Ohashi}, {Bourke}, {Qi}, \& {Zhang}}]{Kuan04}
{Kuan}, Y., {Huang}, H., {Charnley}, S.~B., {et~al.} 2004, \apjl, 616, L27

\bibitem[{{Maret} {et~al.}(2005){Maret}, {Ceccarelli}, {Tielens}, {Caux},
  {Lefloch}, {Faure}, {Castets}, \& {Flower}}]{Maret05}
{Maret}, S., {Ceccarelli}, C., {Tielens}, A.~G.~G.~M., {et~al.} 2005, \aap,
  442, 527

\bibitem[{{M{\"u}ller} {et~al.}(2001){M{\"u}ller}, {Thorwirth}, {Roth}, \&
  {Winnewisser}}]{Muller01}
{M{\"u}ller}, H.~S.~P., {Thorwirth}, S., {Roth}, D.~A., \& {Winnewisser}, G.
  2001, \aap, 370, L49

\bibitem[{{Nomura} \& {Millar}(2004)}]{Nomura04}
{Nomura}, H. \& {Millar}, T.~J. 2004, \aap, 414, 409

\bibitem[{{{\"O}berg} {et~al.}(2009{\natexlab{a}}){{\"O}berg}, {Bottinelli}, \&
  {van Dishoeck}}]{Oberg09a}
{{\"O}berg}, K.~I., {Bottinelli}, S., \& {van Dishoeck}, E.~F.
  2009{\natexlab{a}}, \aap, 494, L13

\bibitem[{{{\"O}berg} {et~al.}(2009{\natexlab{b}}){{\"O}berg}, {Garrod}, {van
  Dishoeck}, \& {Linnartz}}]{Oberg09d}
{{\"O}berg}, K.~I., {Garrod}, R.~T., {van Dishoeck}, E.~F., \& {Linnartz}, H.
  2009{\natexlab{b}}, \aap, 504, 891

\bibitem[{{Pravdo} {et~al.}(2001){Pravdo}, {Feigelson}, {Garmire}, {Maeda},
  {Tsuboi}, \& {Bally}}]{Pravdo01}
{Pravdo}, S.~H., {Feigelson}, E.~D., {Garmire}, G., {et~al.} 2001, \nat, 413,
  708

\bibitem[{{Reipurth} \& {Bally}(2001)}]{Reipurth01}
{Reipurth}, B. \& {Bally}, J. 2001, \araa, 39, 403

\bibitem[{{Schuster} {et~al.}(2004){Schuster}, {Boucher}, {Brunswig}, {Carter},
  {Chenu}, {Foullieux}, {Greve}, {John}, {Lazareff}, {Navarro}, {Perrigouard},
  {Pollet}, {Sievers}, {Thum}, \& {Wiesemeyer}}]{Schuster04}
{Schuster}, K., {Boucher}, C., {Brunswig}, W., {et~al.} 2004, \aap, 423, 1171

\bibitem[{{Shen} {et~al.}(2004){Shen}, {Greenberg}, {Schutte}, \& {van
  Dishoeck}}]{Shen04}
{Shen}, C.~J., {Greenberg}, J.~M., {Schutte}, W.~A., \& {van Dishoeck}, E.~F.
  2004, \aap, 415, 203

\bibitem[{{Spaans} {et~al.}(1995){Spaans}, {Hogerheijde}, {Mundy}, \& {van
  Dishoeck}}]{Spaans95}
{Spaans}, M., {Hogerheijde}, M.~R., {Mundy}, L.~G., \& {van Dishoeck}, E.~F.
  1995, \apjl, 455, L167+

\bibitem[{{van Dishoeck} {et~al.}(1995){van Dishoeck}, {Blake}, {Jansen}, \&
  {Groesbeck}}]{vanDishoeck95}
{van Dishoeck}, E.~F., {Blake}, G.~A., {Jansen}, D.~J., \& {Groesbeck}, T.~D.
  1995, \apj, 447, 760

\bibitem[{{Walawender} {et~al.}(2009){Walawender}, {Reipurth}, \&
  {Bally}}]{Walawender09}
{Walawender}, J., {Reipurth}, B., \& {Bally}, J. 2009, \aj, 137, 3254

\bibitem[{{Watanabe} {et~al.}(2003){Watanabe}, {Shiraki}, \&
  {Kouchi}}]{Watanabe03}
{Watanabe}, N., {Shiraki}, T., \& {Kouchi}, A. 2003, \apjl, 588, L121

\end{thebibliography}
\end{document}